\documentclass[%
reprint,
superscriptaddress,
amsmath,amssymb,
aps
]{revtex4-2}

\usepackage{hyperref}
\hypersetup{colorlinks,allcolors=blue}
\usepackage{graphicx}
\usepackage{bm}
\usepackage[all]{hypcap} 
\usepackage{color}
\usepackage{siunitx}
\usepackage{soul}
\usepackage{tabularx}

\begin{document}

\title{Doping dependence of 2-spinon excitations in the doped 1D cuprate Ba$_2$CuO$_{3+\delta}$}

\author{Jiarui Li}
\thanks{These two authors contributed equally.}
\affiliation{Stanford Institute for Materials and Energy Sciences, SLAC National Accelerator Laboratory, Menlo Park, CA 94025, USA}

\author{Daniel Jost}
\thanks{These two authors contributed equally.}
\affiliation{Stanford Institute for Materials and Energy Sciences, SLAC National Accelerator Laboratory, Menlo Park, CA 94025, USA}

\author{Ta Tang}
\affiliation{Stanford Institute for Materials and Energy Sciences, SLAC National Accelerator Laboratory, Menlo Park, CA 94025, USA}
\affiliation{Department of Applied Physics, Stanford University, Stanford, CA 94305, USA}

\author{Ruohan Wang}
\affiliation{Stanford Institute for Materials and Energy Sciences, SLAC National Accelerator Laboratory, Menlo Park, CA 94025, USA}

\author{Yong Zhong}
\affiliation{Stanford Institute for Materials and Energy Sciences, SLAC National Accelerator Laboratory, Menlo Park, CA 94025, USA}

\author{Zhuoyu Chen}
\affiliation{Department of Physics, Southern University of Science and Technology, Shenzhen 518055, China}

\author{Mirian Garcia-Fernandez}
\affiliation{Diamond Light Source, Harwell Campus, Didcot, Oxfordshire OX11 0DE, United Kingdom}

\author{Jonathan Pelliciari}
\affiliation{National Synchrotron Light Source II, Brookhaven National Laboratory, Upton, NY, 11973, USA}

\author{Valentina Bisogni}
\affiliation{National Synchrotron Light Source II, Brookhaven National Laboratory, Upton, NY, 11973, USA}

\author{Brian Moritz}
\affiliation{Stanford Institute for Materials and Energy Sciences, SLAC National Accelerator Laboratory, Menlo Park, CA 94025, USA}

\author{Kejin Zhou}
\affiliation{Diamond Light Source, Harwell Campus, Didcot, Oxfordshire OX11 0DE, United Kingdom}

\author{Yao Wang}
\affiliation{Department of Chemistry, Emory University, Atlanta, GA 30322, USA}
\affiliation{Department of Physics and Astronomy, Clemson University, Clemson, SC 29634, USA}

\author{Thomas P. Devereaux}
\affiliation{Stanford Institute for Materials and Energy Sciences, SLAC National Accelerator Laboratory, Menlo Park, CA 94025, USA}
\affiliation{Department of Materials Science and Engineering, Stanford University, Stanford, CA 94305, USA}
\affiliation{Geballe Laboratory for Advanced Materials, Stanford University, Stanford, CA 94305, USA}

\author{Wei-Sheng Lee}
\email[Corresponding author: ]{leews@stanford.edu}
\affiliation{Stanford Institute for Materials and Energy Sciences, SLAC National Accelerator Laboratory, Menlo Park, CA 94025, USA}

\author{Zhi-Xun Shen}
\email[Corresponding author: ]{zxshen@stanford.edu}
\affiliation{Stanford Institute for Materials and Energy Sciences, SLAC National Accelerator Laboratory, Menlo Park, CA 94025, USA}
\affiliation{Geballe Laboratory for Advanced Materials, Stanford University, Stanford, CA 94305, USA}
\affiliation{Department of Physics, Stanford University, Stanford, 94305, CA, USA}
\affiliation{Department of Applied Physics, Stanford University, Stanford, 94305, CA, USA}

\date{\today}
\begin{abstract}
Recent photoemission experiments on the quasi-one-dimensional Ba-based cuprates suggest that doped holes experience an attractive potential not captured using the simple Hubbard model. This observation has garnered significant attention due to its potential relevance to Cooper pair formation in high-$T_c$ cuprate superconductors. To scrutinize this assertion, we examined signatures of such an attractive potential in doped 1D cuprates Ba$_2$CuO$_{3+\delta}$ by measuring the dispersion of the 2-spinon excitations using Cu $L_3$-edge resonant inelastic X-ray scattering (RIXS). Upon doping, the 2-spinon excitations appear to weaken, with a shift of the minimal position corresponding to the nesting vector of the Fermi points, $q_F$. Notably, we find that the energy scale of the 2-spinons near the Brillouin zone boundary is substantially softened compared to that predicted by the Hubbard model in one-dimension. Such a discrepancy implies missing ingredients, which lends support for the presence of an additional attractive potential between holes.
\end{abstract}
\maketitle

\begin{figure*}
	\centering
	\includegraphics[width = 2 \columnwidth]{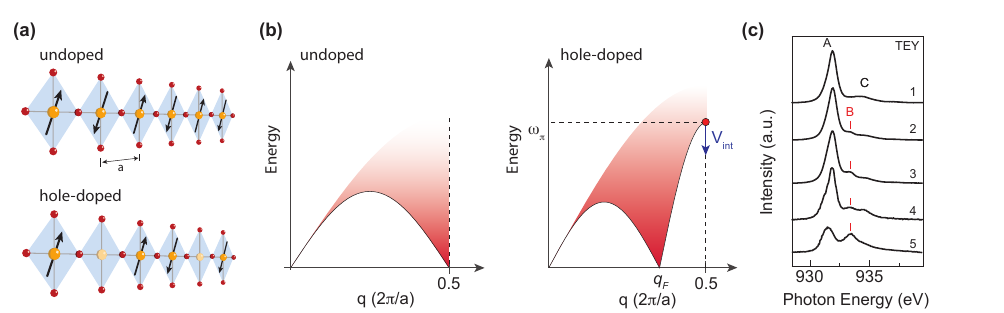}
	\caption{\label{Intro}
	(a) Schematic of a corner-sharing quasi-1D cuprate as an antiferromagnetic spin chain with the lattice constant a. (b) Illustration of 2-spinon excitations for the undoped and doped 1D spin chain, respectively. $\omega_{\pi}$ denotes the 2-spinon energy at $q = 0.5$, which is predicted to soften in the presence of an attractive potential $V_{int}$. (c) XAS obtained by measuring total electron yield (TEY) across the Cu $L_3$-edge of the five samples prepared for this study. A, B and C denote the spectral features around 931.9 eV, 933.4 eV, and 934.5 eV, respectively.
    }
\end{figure*}

Corner-sharing quasi-one-dimensional (1D) cuprates consist of chains of CuO$_2$ plaquettes which are connected via corner sharing oxygen atoms (Fig. \ref{Intro}(a)). They can be modeled as a 1D spin chain with a strong antiferromagnetic (AFM) super-exchange interaction. Therefore, these compounds contain the essence of electronic correlations exhibited in the quasi-two dimensional cuprate superconductors, but with a simpler geometry. Experiments have revealed a wealth of many-body phenomena, including magnetic order, spin-charge separation, and fractionalization, along with associated collective charge, spin, and orbital excitations \cite{Kim1996, Fujisawa1999, Kim2006, Schlappa_Nature_2012, Schlappa_NatComm_2018, Chen_Science_2021}. Such phenomena have provided valuable insight into the effects of strong electronic correlations and the mystery of high-$T_c$ superconductivity.

Recent angle-resolved photoemission spectroscopy (ARPES) experiments on doped 1D cuprates, specifically Ba$_2$CuO$_{3+\delta}$, revealed a doping dependence to the holon and spinon bands, not captured by the simple Hubbard model~\cite{Chen_Science_2021}. An additional strong, extended, and most importantly {\it attractive} interaction was required to model the data effectively. Subsequent theories suggest that incorporating such a potential into the Mott-Hubbard model could enhance superconducting pairing \cite{Qu2022, Zhang2022, Jiang2022, Zhou2023}. Therefore, the results of these experiments and the proposed attractive couplings could have profound implications for understanding the pairing mechanism in cuprate superconductors.

Given the importance of this hypothesis, additional experiments are necessary for validation. Theories predict that such an interaction should also leave fingerprints in the bosonic excitation channels, particularly, the dynamic spin structure factor $S(q,\omega)$ \cite{Shen_arxiv_2024, Tang_arxiv_2024, tohyama2024}. The doping evolution of $S(q,\omega)$ in the simple Hubbard model is sketched in Fig. \ref{Intro}(b). When the system is undoped, $S(q,\omega)$ exhibits a typical two-spinon continuum, a characteristic feature of 1D quantum spin chains due to electron fractionalization. The peak position of $S(q,\omega)$ follows the lower edge of the continuum, which emanates from the zone center, reaching a maximum at $q = 0.25$, then terminating at the Brillouin zone boundary ($q = 0.5$), where the excitation becomes gapless. Upon hole doping, the momentum where the 2-spinon excitations become gapless shifts to $q_F$, corresponding to the nesting wavevector of the Fermi points in the spinon bands. At the zone boundary $q = 0.5$, the excitations become gapped with an energy of $\omega_\pi$. In the presence of an attractive potential $V_{int}$, $\omega_\pi$ will soften and the spectral weight should spread out in energy and momentum. The soften of $\omega_\pi$ can serve as an estimate for the strength of $V_{int}$.

In this letter, we measure the doping dependence of 2-spinon excitations in Ba$_2$CuO$_{3+\delta}$ thin films using Cu $L_3$-edge resonant inelastic x-ray scattering (RIXS). In other undoped one-dimensional cuprates, Cu $L$-edge RIXS captures 2-spinon excitations, with a peak position that tracks the lower boundary of the 2-spinon continuum \cite{Schlappa_Nature_2012,Bisogni2014,Forte2011}, laying the foundations for the doping dependence study reported here. The higher light-matter scattering cross-section and element specificity make RIXS an ideal tool for probing Ba$_2$CuO$_{3+\delta}$ films. Upon doping, we find that the 2-spinon excitations broaden, accompanied by a systematic shift of the gapless momentum $q_F$. We also find that the energy scale of the 2-spinons near the zone boundary, $\omega_{\pi}$, is substantially lower than predicted by the one-dimensional Hubbard model. Such a discrepancy implies missing ingredients in the Hubbard model and supports the presence of an additional attractive potential between holes.

Ba$_2$CuO$_{3+\delta}$ thin films were synthesized through ozone-reactive molecular beam epitaxy (MBE). Ba and Cu sources were co-evaporated onto 001 oriented SrTiO$_{3}$ substrates with TiO$_{2}$ termination. Doping was controlled via the temperature at which the ozone supply was shut off. More hole doping is expected for the sample grown with a lower ozone shut-off temperature. The thickness of samples 1-3 is 50 nm, with ozone shut-off temperatures of 350°C, 270°C, and 180°C followed by 10 minutes of UHV annealing. We found it very challenging to achieve higher doping concentrations in 50 nm thick films, requiring thinner films for homogeneous doping across samples (samples 4 and 5 with a thickness of 10 nm). Sample 4 had an ozone shut-off temperature of 440°C, followed by 10 minutes of UHV annealing. The sample with the highest doping concentration (sample 5) was prepared by annealing in ozone at growth temperature (570°C) for 5 minutes and cooling down in ozone until the sample temperature reached 100°C. More details on sample synthesis can be found in the supplementary material \cite{Sup}.

Cu $L_3$-edge XAS and RIXS experiments were performed at the 2-ID SIX of NSLS-II for samples 1-3 and at I21 of Diamond Light Source for samples 4 and 5. XAS data were obtained by measuring total electron yield at a grazing incident angle of 20°. The x-ray polarization was parallel to the scattering plane (\textit{i.e.}, $\pi$-geometry). RIXS data were taken with the spectrometer angle $2\theta$ set at $\sim$150° also using a $\pi$-polarization geometry, which is known to be more sensitive to magnetic excitations. As magnetic excitations in 1D systems only disperse along the chain direction, all momentum-dependent data have been denoted as a function of the projected in-chain momentum transfer $q$ along the chain direction using the reciprocal lattice unit $2\pi/a$ ($a$ =3.90 \AA). A detailed schematic of the experimental geometry is show in the supplementary Fig. S1 \cite{Sup}. The energy resolution of the RIXS data was 32 meV for samples 1-3 and 42 meV for samples 4 and 5. All XAS and RIXS measurements were conducted at approximately 30 K. \color{black} The intensity of all RIXS spectra is normalized to the integrated intensity of the $dd$ excitations. The maxima of 2-spinon excitations were extracted by fitting to a Gaussian function (see supplementary material \cite{Sup}). \color{black}

\begin{figure*}
	\centering
	\includegraphics[width = 2 \columnwidth]{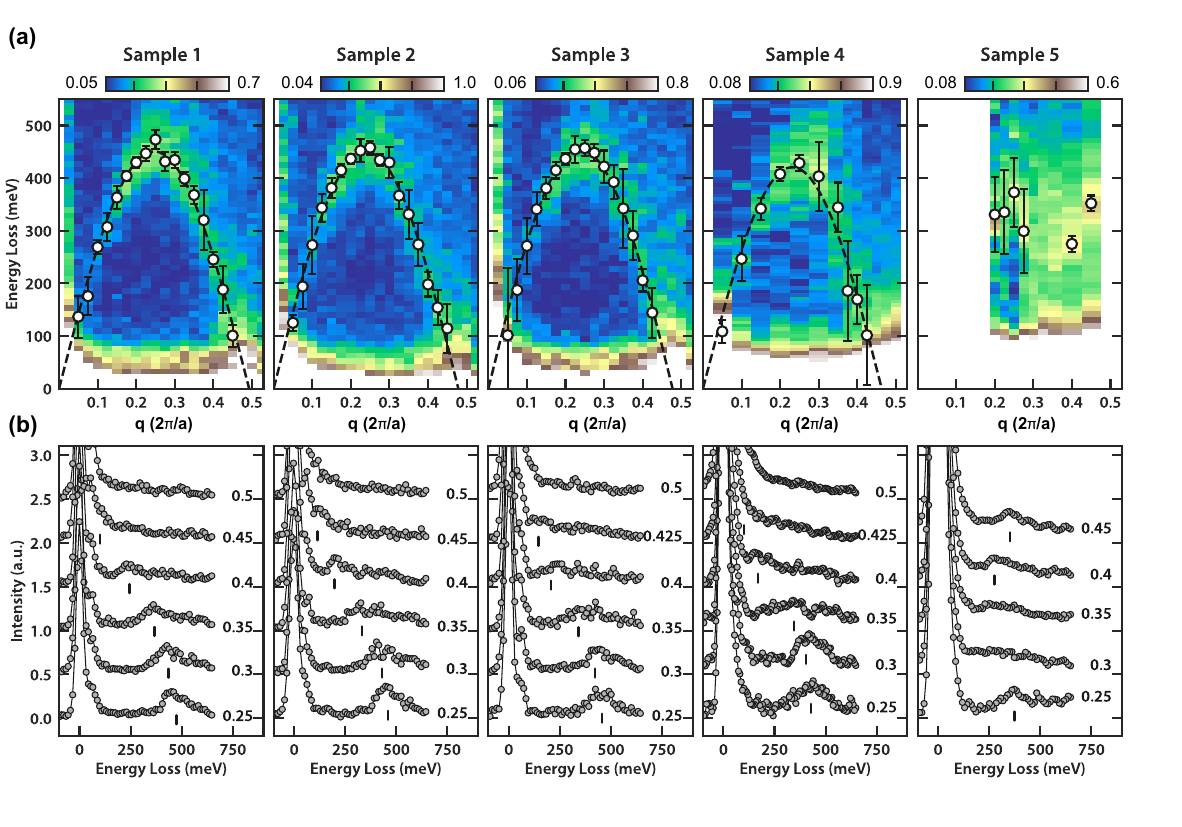}
	\caption{\label{2Spinon}
	(a) RIXS intensity map as a function of energy loss and momentum projected along the chain direction for Samples 1-5. The markers indicate the peak position of the 2-spinon spectra. \color{black} The error bars are the 90\% confidence intervals of the fitted peak position. \color{black} Black dashed lines indicate the fit for each sample's theoretical lower bounds of the 2-spinon dispersion. (b) Stack plots of RIXS energy loss spectra at representative momenta for each sample. Vertical lines mark the peak positions of the 2-spinon excitations.
    }
\end{figure*}

We first verify that holes were introduced into the compounds by our synthesis method. Figure \ref{Intro}(d) presents the X-ray absorption spectra (XAS) across the Cu $L_3$-edge. In sample 1, which is the least hole-doped, the XAS is essentially identical to those measured in the single-crystal undoped 1D cuprate, Sr$_2$CuO$_3$~\cite{Schlappa_Nature_2012}. Specifically, the XAS consists of a main peak, A, at 931.9 eV, and a weak hump, C, extending up to 934.5 eV. Locally, the ground state of the undoped compound is primarily of 3$d^9$ and 3$d^{10}\underline{L}$ character ($\underline{L}$ denotes the ligand character of the ground state from the oxygen 2$p$ orbital contributions). The main peak A corresponds to transitions into the 3$d^{10}$ final states, and the weak high-energy hump C has been attributed to the transition to Cu 3$d_{3z^2-r^2}$ orbitals, which become partially unoccupied due to hybridization with the Cu 4$s$ states \cite{Fink_1994, Neudert2000, Salluzzo2007}.

For samples 2 to 5, where more hole doping is expected, a new peak (denoted B) located at 933.4 eV emerges with increasing spectral weight with nominally increasing hole concentration, reminiscent of the hole-induced high energy shoulder next to the main XAS peak observed in 2D cuprates. Here, the feature is more pronounced, likely due to poor screening in one dimension \cite{Chen_PRL_1992, Veenendaal_1994, Ghiringhelli_PRB_2007}. The ground state of the hole-doped Cu sites should possess a dominant 3$d^9\underline{L}$ character (doped holes on oxygen); therefore, peak B can be attributed to the transition locally to the 3$d^{10}\underline{L}$ final state. Thus, the growth of spectral weight in peak B demonstrates a systematic increase of hole concentration in our series of Ba$_2$CuO$_3$ samples \cite{Fink_1994, Neudert2000, Salluzzo2007, Wang_jes_2014}.

By tuning the photon energy to the main peak of the XAS, RIXS can map out the energy-momentum dispersion of the 2-spinon excitations. Figure \ref{2Spinon}(a) summarizes the RIXS intensity map as a function of projected momentum along the chain direction. RIXS spectra at representative momenta are displayed in the lower panel (Fig. \ref{2Spinon} (b)). In the least doped sample 1, we observe a branch of excitations whose peak positions reach a maximum at $q = 0.25$ with an energy of approximately 450 meV. This merges into the elastic peak near the zone boundary ($q = 0.5$), indicating that the excitations become gapless. This dispersion is expected for the lower edge of the 2-spinon continuum in the undoped 1D antiferromagnetic spin chain (see also Fig. \ref{Intro}(b)), with an energy scale as in single crystal Sr$_2$CuO$_3$ \cite{Schlappa_Nature_2012}. In addition, like single crystal Sr$_2$CuO$_3$\cite{Schlappa_Nature_2012}, the $dd$ excitations exhibit a substantial energy-momentum dispersion, which has been attributed to the ``spin-orbital" separation (see supplementary Fig. S2 ~\cite{Sup}).

Upon hole doping, as shown in Fig. \ref{2Spinon}, although the overall features of the 2-spinon dispersion remain similar, the excitations become broader in energy-momentum space, and the maximal 2-spinon energy decreases at the lower edge of the contiuum near $q = 0.25$. Importantly, the momentum $q_F$ where the 2-spinon excitation becomes gapless shifts from $q = 0.5$ to smaller values with increasing hole doping. In the most heavily hole-doped sample 5, as shown in Fig. \ref{2Spinon}(b), the 2-spinon continuum becomes gapless at a momentum between $q = 0.3$ and $0.35$, where the peak merges with elastic peak and becomes invisible. Beyond this momentum, the peak becomes observable again and increases in energy as it approaches the zone boundary at $q = 0.5$. Taken together, we have observed an expected systematic shift of $q_F$ with increasing hole doping.

$q_F$ should correspond to the nesting wave-vector of the Fermi points in the spinon band with the corresponding doping $x$, which can be expressed as $q_F = (1-x)/2$ in reciprocal lattice units ($2\pi/a$). The doping level of each sample can be estimated more accurately by determining the $q_F$ position in the 2-spinon dispersion. We fit the dispersion $\omega_s(q)$ to 
\begin{equation}
\omega_s(q) = \omega_o \cdot \sin(2\pi\frac{q}{(1-x)}),
\end{equation}
where $\omega_o$ denotes the maximal energy of the 2-spinon continuum near $q=0.25$. For sample 5, the fitting procedure was not applied due to insufficient data, where instead $q_F$ has been determined by the peak positions of the momentum distribution curves of the 2-spinon intensity at constant energies \cite{Sup}. The extracted $\omega_o$ and the doping concentration $x$ are summarized in Table \ref{doping}

\begin{table}[h]
    \centering
    \caption{Fitted parameters for 2-spinon excitations}
    \label{doping}
    \begin{tabularx}{1.018\columnwidth}{|c|c|c|c|c|c|}
        \hline
        Sample & 1 & 2 & 3 & 4 & 5 \\ 
        \hline
        $\omega_{o}$ (meV) & 446 $\pm$ 4 & 447 $\pm$ 4 & 453 $\pm$ 5 & 417 $\pm$ 13 & 360 $\pm$ 19 \\ 
        \hline
        $x$ (\%) & 1.3 $\pm$ 0.6 & 4.2 $\pm$ 0.6 & 3.9 $\pm$ 0.8 & 8.1 $\pm$ 1.6 & 33 $\pm$ 4 \\ 
        \hline
    \end{tabularx}
\end{table}

\begin{figure}
	\centering
	\includegraphics[width = 1.03 \columnwidth]{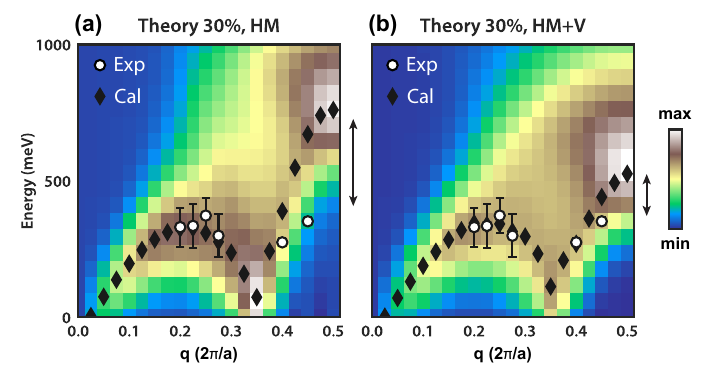}
	\caption{\label{Theory}
	Calculated 2-spinon excitations in a 30 \% hole doped 1D spin chain using (a) Hubbard model and (b) extended-Hubbard model with an effective nearest-neighbor attractive interaction $V = -1t$. \color{black} The calculated peak positions (black markers) and the 2-spinon peak positions in sample 5 (open markers, shown also in Fig. \ref{2Spinon}(a)) \color{black} are superimposed for comparison. \color{black} The discrepancy between experiment and calculated 2-spinon excitation energy at the zone boundary is indicated by vertical arrows, as a guide to the eye.\color{black} The calculation were reproduced from Ref. \cite{Tang_arxiv_2024}.}
\end{figure}

With these estimated doping concentrations, our data can be compared to the 1D Hubbard-like model simulations with similar hole concentrations. Figure \ref{Theory}(a) shows the calculated 2-spinon excitations with a hole doping of $x = 30\%$ using a simple Hubbard model (HM), compared to the peak positions extracted from sample 5 (white markers). The energy scale of the calculation was determined by normalizing the 2-spinon peak energy at $q=0.25$ in undoped (x = 0\%) to that in our least doped sample 1, setting an overall energy scale of the hopping integral in the Hubbard model $t$ = 0.7 eV, similar to scaling in Ref. \cite{Chen_Science_2021}. More details about the calculation are described elsewhere \cite{Tang_arxiv_2024,Sup}. While the calculated 2-spinon excitations capture the essence of the measured dispersion for $q<q_F$, the Hubbard model significantly overestimates the energy scale near the zone boundary compared to the experimental data \color{black} (see Fig. \ref{Theory}(a) and supplementary material)\color{black}. This discrepancy indicates that the Hubbard model alone cannot fully account for the doping dependence of the 2-spinon excitations in Ba$_2$CuO$_{3+\delta}$.

Theory has shown that by adding an attractive potential in the doped 1D Hubbard model, the characteristic energy scale $\omega_{\pi}$ decreases, and the spectral weight broadens in energy and momentum~\cite{Tang_arxiv_2024, Shen_arxiv_2024, tohyama2024}. Figure \ref{Theory}(b) shows the calculated 2-spinon excitations with an effective attractive interaction $V = -1.0t$, showing a better agreement with the data than that predicted in the simple Hubbard model. While a slightly larger value of $V$ might better fit the experimental data, such fine-tuning may not be as meaningful due to the uncertainty in the data and the doping concentration of sample. Nevertheless, this is a substantial attractive interaction and is consistent with the value previously determined by ARPES~\cite{Chen_Science_2021}.

It might be interesting to compare with recent work on another quasi-1D antiferromagnetic (AFM) spin chain, YbAlO$_3$. The spinon band filling was effectively tuned by a magnetic field, and the evolution of the 2-spinon continuum was measured by inelastic neutron scattering~\cite{Nikitin2021}. Notably, at a magnetic field of 0.8 T, where $q_F \sim 0.3~(2\pi/a)$, the $\omega{_\pi}$ is approximately two to three times higher than $\omega_{o}$. This is similar to that in the simple Hubbard model with a comparable $q_F$ due to hole doping (see, for example, Fig. \ref{Theory}(a)). This agreement is in stark contrast with our experimental observations on Ba$2$CuO$_{3+\delta}$, highlighting the presence of an attractive potential in quasi-1D cuprates. We suspect that the origin of the unique coupling may stem from charge transfer between the Cu 3$d$ orbitals and O ligand, which appears to be absent in the YbAlO$_3$. Note that charge transfer allows oxygen phonons to more strongly couple with the electrons, which could give rise to such an attractive potential \cite{Wang_2021}.

\color{black} 

In conclusion, our results provide further support for the existence of a long-range attractive potential in quasi-1D cuprates that is absent in the simple Hubbard model. While it has been suggested that this potential may arise from electron-phonon coupling \cite{Wang_2021,Tang2023,Tang_arxiv_2024, Wang2024}, verifying this intriguing hypothesis is beyond the scope of this manuscript. A complimentary study using RIXS at the O $K$-edge with greater sensitivity to electron-{\it oxygen} phonon coupling may shed additional light on this subject~\cite{Devereaux2016,Jost2024}. Further investigations into the origin of this potential and how it manifests in higher dimensions could be important for understanding and controlling the properties of high-$T_c$ cuprates.

\section{ACKNOWLEDGEMENTS}

This work was supported by the U.S. Department of Energy, Office of Basic Energy Sciences, Division of Materials Sciences and Engineering, under Contract No. DE-AC02-76SF00515. J.L. acknowledges the support of the Department of Energy, Laboratory Directed Research and Development program at SLAC National Accelerator Laboratory, under contract DE-AC02-76SF00515. D.J. gratefully acknowledges funding by the Alexander-von-Humboldt foundation. Part of data were taken at the I21 RIXS beamline of Diamond Light Source (UK) using the RIXS spectrometer designed, built, and owned by Diamond Light Source. We thank Diamond Light Source for providing beamtime under proposal ID MM32900.
This research used beamline 2-ID of NSLS-II, a US DOE Office of Science User Facility operated for the DOE Office of Science by Brookhaven National Laboratory under contract no. DE-SC0012704.
Y.W. acknowledges support from the Air Force Office of Scientific Research Young Investigator Program under grant FA9550-23-1-0153. The computational results utilized the resources of the National Energy Research Scientific Computing Center (NERSC), a Department of Energy Office of Science User Facility using NERSC award BES-ERCAP0023181. Some of the computing for this project was performed on the Sherlock cluster. We would like to thank Stanford University and the Stanford Research Computing Center for providing computational resources and support that contributed to these research results.

\bibliography{reference}

\begin{thebibliography}{29}%
\makeatletter
\providecommand \@ifxundefined [1]{%
 \@ifx{#1\undefined}
}%
\providecommand \@ifnum [1]{%
 \ifnum #1\expandafter \@firstoftwo
 \else \expandafter \@secondoftwo
 \fi
}%
\providecommand \@ifx [1]{%
 \ifx #1\expandafter \@firstoftwo
 \else \expandafter \@secondoftwo
 \fi
}%
\providecommand \natexlab [1]{#1}%
\providecommand \enquote  [1]{``#1''}%
\providecommand \bibnamefont  [1]{#1}%
\providecommand \bibfnamefont [1]{#1}%
\providecommand \citenamefont [1]{#1}%
\providecommand \href@noop [0]{\@secondoftwo}%
\providecommand \href [0]{\begingroup \@sanitize@url \@href}%
\providecommand \@href[1]{\@@startlink{#1}\@@href}%
\providecommand \@@href[1]{\endgroup#1\@@endlink}%
\providecommand \@sanitize@url [0]{\catcode `\\12\catcode `\$12\catcode `\&12\catcode `\#12\catcode `\^12\catcode `\_12\catcode `\%12\relax}%
\providecommand \@@startlink[1]{}%
\providecommand \@@endlink[0]{}%
\providecommand \url  [0]{\begingroup\@sanitize@url \@url }%
\providecommand \@url [1]{\endgroup\@href {#1}{\urlprefix }}%
\providecommand \urlprefix  [0]{URL }%
\providecommand \Eprint [0]{\href }%
\providecommand \doibase [0]{https://doi.org/}%
\providecommand \selectlanguage [0]{\@gobble}%
\providecommand \bibinfo  [0]{\@secondoftwo}%
\providecommand \bibfield  [0]{\@secondoftwo}%
\providecommand \translation [1]{[#1]}%
\providecommand \BibitemOpen [0]{}%
\providecommand \bibitemStop [0]{}%
\providecommand \bibitemNoStop [0]{.\EOS\space}%
\providecommand \EOS [0]{\spacefactor3000\relax}%
\providecommand \BibitemShut  [1]{\csname bibitem#1\endcsname}%
\let\auto@bib@innerbib\@empty
\bibitem [{\citenamefont {Kim}\ \emph {et~al.}(1996)\citenamefont {Kim}, \citenamefont {Matsuura}, \citenamefont {Shen}, \citenamefont {Motoyama}, \citenamefont {Eisaki}, \citenamefont {Uchida}, \citenamefont {Tohyama},\ and\ \citenamefont {Maekawa}}]{Kim1996}%
  \BibitemOpen
  \bibfield  {author} {\bibinfo {author} {\bibfnamefont {C.}~\bibnamefont {Kim}}, \bibinfo {author} {\bibfnamefont {A.~Y.}\ \bibnamefont {Matsuura}}, \bibinfo {author} {\bibfnamefont {Z.-X.}\ \bibnamefont {Shen}}, \bibinfo {author} {\bibfnamefont {N.}~\bibnamefont {Motoyama}}, \bibinfo {author} {\bibfnamefont {H.}~\bibnamefont {Eisaki}}, \bibinfo {author} {\bibfnamefont {S.}~\bibnamefont {Uchida}}, \bibinfo {author} {\bibfnamefont {T.}~\bibnamefont {Tohyama}},\ and\ \bibinfo {author} {\bibfnamefont {S.}~\bibnamefont {Maekawa}},\ }\bibfield  {title} {\bibinfo {title} {{Observation of Spin-Charge Separation in One-Dimensional SrCu${\mathrm{O}}_{2}$}},\ }\href {https://doi.org/10.1103/PhysRevLett.77.4054} {\bibfield  {journal} {\bibinfo  {journal} {Phys. Rev. Lett.}\ }\textbf {\bibinfo {volume} {77}},\ \bibinfo {pages} {4054} (\bibinfo {year} {1996})}\BibitemShut {NoStop}%
\bibitem [{\citenamefont {Fujisawa}\ \emph {et~al.}(1999)\citenamefont {Fujisawa}, \citenamefont {Yokoya}, \citenamefont {Takahashi}, \citenamefont {Miyasaka}, \citenamefont {Kibune},\ and\ \citenamefont {Takagi}}]{Fujisawa1999}%
  \BibitemOpen
  \bibfield  {author} {\bibinfo {author} {\bibfnamefont {H.}~\bibnamefont {Fujisawa}}, \bibinfo {author} {\bibfnamefont {T.}~\bibnamefont {Yokoya}}, \bibinfo {author} {\bibfnamefont {T.}~\bibnamefont {Takahashi}}, \bibinfo {author} {\bibfnamefont {S.}~\bibnamefont {Miyasaka}}, \bibinfo {author} {\bibfnamefont {M.}~\bibnamefont {Kibune}},\ and\ \bibinfo {author} {\bibfnamefont {H.}~\bibnamefont {Takagi}},\ }\bibfield  {title} {\bibinfo {title} {{Angle-resolved photoemission study of ${\mathrm{Sr}}_{2}{\mathrm{CuO}}_{3}$}},\ }\href {https://doi.org/10.1103/PhysRevB.59.7358} {\bibfield  {journal} {\bibinfo  {journal} {Phys. Rev. B}\ }\textbf {\bibinfo {volume} {59}},\ \bibinfo {pages} {7358} (\bibinfo {year} {1999})}\BibitemShut {NoStop}%
\bibitem [{\citenamefont {Kim}\ \emph {et~al.}(2006)\citenamefont {Kim}, \citenamefont {Koh}, \citenamefont {Rotenberg}, \citenamefont {Oh}, \citenamefont {Eisaki}, \citenamefont {Motoyama}, \citenamefont {Uchida}, \citenamefont {Tohyama}, \citenamefont {Maekawa}, \citenamefont {Shen},\ and\ \citenamefont {Kim}}]{Kim2006}%
  \BibitemOpen
  \bibfield  {author} {\bibinfo {author} {\bibfnamefont {B.~J.}\ \bibnamefont {Kim}}, \bibinfo {author} {\bibfnamefont {H.}~\bibnamefont {Koh}}, \bibinfo {author} {\bibfnamefont {E.}~\bibnamefont {Rotenberg}}, \bibinfo {author} {\bibfnamefont {S.~J.}\ \bibnamefont {Oh}}, \bibinfo {author} {\bibfnamefont {H.}~\bibnamefont {Eisaki}}, \bibinfo {author} {\bibfnamefont {N.}~\bibnamefont {Motoyama}}, \bibinfo {author} {\bibfnamefont {S.}~\bibnamefont {Uchida}}, \bibinfo {author} {\bibfnamefont {T.}~\bibnamefont {Tohyama}}, \bibinfo {author} {\bibfnamefont {S.}~\bibnamefont {Maekawa}}, \bibinfo {author} {\bibfnamefont {Z.~X.}\ \bibnamefont {Shen}},\ and\ \bibinfo {author} {\bibfnamefont {C.}~\bibnamefont {Kim}},\ }\bibfield  {title} {\bibinfo {title} {{Distinct spinon and holon dispersions in photoemission spectral functions from one-dimensional SrCuO$_2$}},\ }\href {https://doi.org/10.1038/nphys316} {\bibfield  {journal} {\bibinfo  {journal} {Nature Physics}\ }\textbf {\bibinfo {volume} {2}},\ \bibinfo {pages}
  {397} (\bibinfo {year} {2006})}\BibitemShut {NoStop}%
\bibitem [{\citenamefont {Schlappa}\ \emph {et~al.}(2012)\citenamefont {Schlappa}, \citenamefont {Wohlfeld}, \citenamefont {Zhou}, \citenamefont {Mourigal}, \citenamefont {Haverkort}, \citenamefont {Strocov}, \citenamefont {Hozoi}, \citenamefont {Monney}, \citenamefont {Nishimoto}, \citenamefont {Singh}, \citenamefont {Revcolevschi}, \citenamefont {Caux}, \citenamefont {Patthey}, \citenamefont {Rønnow}, \citenamefont {van~den Brink},\ and\ \citenamefont {Schmitt}}]{Schlappa_Nature_2012}%
  \BibitemOpen
  \bibfield  {author} {\bibinfo {author} {\bibfnamefont {J.}~\bibnamefont {Schlappa}}, \bibinfo {author} {\bibfnamefont {K.}~\bibnamefont {Wohlfeld}}, \bibinfo {author} {\bibfnamefont {K.~J.}\ \bibnamefont {Zhou}}, \bibinfo {author} {\bibfnamefont {M.}~\bibnamefont {Mourigal}}, \bibinfo {author} {\bibfnamefont {M.~W.}\ \bibnamefont {Haverkort}}, \bibinfo {author} {\bibfnamefont {V.~N.}\ \bibnamefont {Strocov}}, \bibinfo {author} {\bibfnamefont {L.}~\bibnamefont {Hozoi}}, \bibinfo {author} {\bibfnamefont {C.}~\bibnamefont {Monney}}, \bibinfo {author} {\bibfnamefont {S.}~\bibnamefont {Nishimoto}}, \bibinfo {author} {\bibfnamefont {S.}~\bibnamefont {Singh}}, \bibinfo {author} {\bibfnamefont {A.}~\bibnamefont {Revcolevschi}}, \bibinfo {author} {\bibfnamefont {J.-S.}\ \bibnamefont {Caux}}, \bibinfo {author} {\bibfnamefont {L.}~\bibnamefont {Patthey}}, \bibinfo {author} {\bibfnamefont {H.~M.}\ \bibnamefont {Rønnow}}, \bibinfo {author} {\bibfnamefont {J.}~\bibnamefont {van~den Brink}},\ and\ \bibinfo {author}
  {\bibfnamefont {T.}~\bibnamefont {Schmitt}},\ }\bibfield  {title} {\bibinfo {title} {{Spin–orbital separation in the quasi-one-dimensional Mott insulator Sr$_2$CuO$_3$}},\ }\href {https://doi.org/10.1038/nature10974} {\bibfield  {journal} {\bibinfo  {journal} {Nature}\ }\textbf {\bibinfo {volume} {485}},\ \bibinfo {pages} {82} (\bibinfo {year} {2012})}\BibitemShut {NoStop}%
\bibitem [{\citenamefont {Schlappa}\ \emph {et~al.}(2018)\citenamefont {Schlappa}, \citenamefont {Kumar}, \citenamefont {Zhou}, \citenamefont {Singh}, \citenamefont {Mourigal}, \citenamefont {Strocov}, \citenamefont {Revcolevschi}, \citenamefont {Patthey}, \citenamefont {Rønnow}, \citenamefont {Johnston},\ and\ \citenamefont {Schmitt}}]{Schlappa_NatComm_2018}%
  \BibitemOpen
  \bibfield  {author} {\bibinfo {author} {\bibfnamefont {J.}~\bibnamefont {Schlappa}}, \bibinfo {author} {\bibfnamefont {U.}~\bibnamefont {Kumar}}, \bibinfo {author} {\bibfnamefont {K.~J.}\ \bibnamefont {Zhou}}, \bibinfo {author} {\bibfnamefont {S.}~\bibnamefont {Singh}}, \bibinfo {author} {\bibfnamefont {M.}~\bibnamefont {Mourigal}}, \bibinfo {author} {\bibfnamefont {V.~N.}\ \bibnamefont {Strocov}}, \bibinfo {author} {\bibfnamefont {A.}~\bibnamefont {Revcolevschi}}, \bibinfo {author} {\bibfnamefont {L.}~\bibnamefont {Patthey}}, \bibinfo {author} {\bibfnamefont {H.~M.}\ \bibnamefont {Rønnow}}, \bibinfo {author} {\bibfnamefont {S.}~\bibnamefont {Johnston}},\ and\ \bibinfo {author} {\bibfnamefont {T.}~\bibnamefont {Schmitt}},\ }\bibfield  {title} {\bibinfo {title} {{Probing multi-spinon excitations outside of the two-spinon continuum in the antiferromagnetic spin chain cuprate Sr$_2$CuO$_3$}},\ }\href {https://doi.org/10.1038/s41467-018-07838-y} {\bibfield  {journal} {\bibinfo  {journal} {Nature Communications
  2018 9:1}\ }\textbf {\bibinfo {volume} {9}},\ \bibinfo {pages} {1} (\bibinfo {year} {2018})}\BibitemShut {NoStop}%
\bibitem [{\citenamefont {Chen}\ \emph {et~al.}(2021)\citenamefont {Chen}, \citenamefont {Wang}, \citenamefont {Rebec}, \citenamefont {Jia}, \citenamefont {Hashimoto}, \citenamefont {Lu}, \citenamefont {Moritz}, \citenamefont {Moore}, \citenamefont {Devereaux},\ and\ \citenamefont {Shen}}]{Chen_Science_2021}%
  \BibitemOpen
  \bibfield  {author} {\bibinfo {author} {\bibfnamefont {Z.}~\bibnamefont {Chen}}, \bibinfo {author} {\bibfnamefont {Y.}~\bibnamefont {Wang}}, \bibinfo {author} {\bibfnamefont {S.~N.}\ \bibnamefont {Rebec}}, \bibinfo {author} {\bibfnamefont {T.}~\bibnamefont {Jia}}, \bibinfo {author} {\bibfnamefont {M.}~\bibnamefont {Hashimoto}}, \bibinfo {author} {\bibfnamefont {D.}~\bibnamefont {Lu}}, \bibinfo {author} {\bibfnamefont {B.}~\bibnamefont {Moritz}}, \bibinfo {author} {\bibfnamefont {R.~G.}\ \bibnamefont {Moore}}, \bibinfo {author} {\bibfnamefont {T.~P.}\ \bibnamefont {Devereaux}},\ and\ \bibinfo {author} {\bibfnamefont {Z.-X.}\ \bibnamefont {Shen}},\ }\bibfield  {title} {\bibinfo {title} {Anomalously strong near-neighbor attraction in doped 1d cuprate chains},\ }\href {https://www.science.org} {\bibfield  {journal} {\bibinfo  {journal} {Science}\ }\textbf {\bibinfo {volume} {373}},\ \bibinfo {pages} {1235} (\bibinfo {year} {2021})}\BibitemShut {NoStop}%
\bibitem [{\citenamefont {Qu}\ \emph {et~al.}(2022)\citenamefont {Qu}, \citenamefont {Chen}, \citenamefont {Jiang}, \citenamefont {Wang},\ and\ \citenamefont {Li}}]{Qu2022}%
  \BibitemOpen
  \bibfield  {author} {\bibinfo {author} {\bibfnamefont {D.-W.}\ \bibnamefont {Qu}}, \bibinfo {author} {\bibfnamefont {B.-B.}\ \bibnamefont {Chen}}, \bibinfo {author} {\bibfnamefont {H.-C.}\ \bibnamefont {Jiang}}, \bibinfo {author} {\bibfnamefont {Y.}~\bibnamefont {Wang}},\ and\ \bibinfo {author} {\bibfnamefont {W.}~\bibnamefont {Li}},\ }\bibfield  {title} {\bibinfo {title} {Spin-triplet pairing induced by near-neighbor attraction in the extended hubbard model for cuprate chain},\ }\href {https://doi.org/10.1038/s42005-022-01030-x} {\bibfield  {journal} {\bibinfo  {journal} {Communications Physics}\ }\textbf {\bibinfo {volume} {5}},\ \bibinfo {pages} {257} (\bibinfo {year} {2022})}\BibitemShut {NoStop}%
\bibitem [{\citenamefont {Zhang}\ \emph {et~al.}(2022)\citenamefont {Zhang}, \citenamefont {Guo}, \citenamefont {Mou}, \citenamefont {Chen},\ and\ \citenamefont {Ma}}]{Zhang2022}%
  \BibitemOpen
  \bibfield  {author} {\bibinfo {author} {\bibfnamefont {L.}~\bibnamefont {Zhang}}, \bibinfo {author} {\bibfnamefont {T.}~\bibnamefont {Guo}}, \bibinfo {author} {\bibfnamefont {Y.}~\bibnamefont {Mou}}, \bibinfo {author} {\bibfnamefont {Q.}~\bibnamefont {Chen}},\ and\ \bibinfo {author} {\bibfnamefont {T.}~\bibnamefont {Ma}},\ }\bibfield  {title} {\bibinfo {title} {Enhancement of $d$-wave pairing in the striped phase with nearest neighbor attraction},\ }\href {https://doi.org/10.1103/PhysRevB.105.155154} {\bibfield  {journal} {\bibinfo  {journal} {Phys. Rev. B}\ }\textbf {\bibinfo {volume} {105}},\ \bibinfo {pages} {155154} (\bibinfo {year} {2022})}\BibitemShut {NoStop}%
\bibitem [{\citenamefont {Jiang}(2022)}]{Jiang2022}%
  \BibitemOpen
  \bibfield  {author} {\bibinfo {author} {\bibfnamefont {M.}~\bibnamefont {Jiang}},\ }\bibfield  {title} {\bibinfo {title} {Enhancing $d$-wave superconductivity with nearest-neighbor attraction in the extended hubbard model},\ }\href {https://doi.org/10.1103/PhysRevB.105.024510} {\bibfield  {journal} {\bibinfo  {journal} {Phys. Rev. B}\ }\textbf {\bibinfo {volume} {105}},\ \bibinfo {pages} {024510} (\bibinfo {year} {2022})}\BibitemShut {NoStop}%
\bibitem [{\citenamefont {Zhou}\ \emph {et~al.}(2023)\citenamefont {Zhou}, \citenamefont {Ye}, \citenamefont {Luo}, \citenamefont {Zhao},\ and\ \citenamefont {Chang}}]{Zhou2023}%
  \BibitemOpen
  \bibfield  {author} {\bibinfo {author} {\bibfnamefont {Z.}~\bibnamefont {Zhou}}, \bibinfo {author} {\bibfnamefont {W.}~\bibnamefont {Ye}}, \bibinfo {author} {\bibfnamefont {H.-G.}\ \bibnamefont {Luo}}, \bibinfo {author} {\bibfnamefont {J.}~\bibnamefont {Zhao}},\ and\ \bibinfo {author} {\bibfnamefont {J.}~\bibnamefont {Chang}},\ }\bibfield  {title} {\bibinfo {title} {Robust superconducting correlation against intersite interactions in the extended two-leg hubbard ladder},\ }\href {https://doi.org/10.1103/PhysRevB.108.195136} {\bibfield  {journal} {\bibinfo  {journal} {Phys. Rev. B}\ }\textbf {\bibinfo {volume} {108}},\ \bibinfo {pages} {195136} (\bibinfo {year} {2023})}\BibitemShut {NoStop}%
\bibitem [{\citenamefont {Shen}\ \emph {et~al.}(2024)\citenamefont {Shen}, \citenamefont {Liu}, \citenamefont {Wang},\ and\ \citenamefont {Wang}}]{Shen_arxiv_2024}%
  \BibitemOpen
  \bibfield  {author} {\bibinfo {author} {\bibfnamefont {Z.}~\bibnamefont {Shen}}, \bibinfo {author} {\bibfnamefont {J.}~\bibnamefont {Liu}}, \bibinfo {author} {\bibfnamefont {H.-X.}\ \bibnamefont {Wang}},\ and\ \bibinfo {author} {\bibfnamefont {Y.}~\bibnamefont {Wang}},\ }\href@noop {} {\bibinfo {title} {Signatures of the attractive interaction in spin spectra of 1d cuprate chains}} (\bibinfo {year} {2024}),\ \Eprint {https://arxiv.org/abs/2405.11472} {arXiv:2405.11472} \BibitemShut {NoStop}%
\bibitem [{\citenamefont {Tang}\ \emph {et~al.}(2024)\citenamefont {Tang}, \citenamefont {Jost}, \citenamefont {Moritz},\ and\ \citenamefont {Devereaux}}]{Tang_arxiv_2024}%
  \BibitemOpen
  \bibfield  {author} {\bibinfo {author} {\bibfnamefont {T.}~\bibnamefont {Tang}}, \bibinfo {author} {\bibfnamefont {D.}~\bibnamefont {Jost}}, \bibinfo {author} {\bibfnamefont {B.}~\bibnamefont {Moritz}},\ and\ \bibinfo {author} {\bibfnamefont {T.~P.}\ \bibnamefont {Devereaux}},\ }\href@noop {} {\bibinfo {title} {The influence of extended interactions on spin dynamics in one-dimensional cuprates}} (\bibinfo {year} {2024}),\ \Eprint {https://arxiv.org/abs/2405.11445} {arXiv:2405.11445} \BibitemShut {NoStop}%
\bibitem [{\citenamefont {Tohyama}\ \emph {et~al.}(2024)\citenamefont {Tohyama}, \citenamefont {Maeda}, \citenamefont {Tsutsui}, \citenamefont {Sota},\ and\ \citenamefont {Yunoki}}]{tohyama2024}%
  \BibitemOpen
  \bibfield  {author} {\bibinfo {author} {\bibfnamefont {T.}~\bibnamefont {Tohyama}}, \bibinfo {author} {\bibfnamefont {H.}~\bibnamefont {Maeda}}, \bibinfo {author} {\bibfnamefont {K.}~\bibnamefont {Tsutsui}}, \bibinfo {author} {\bibfnamefont {S.}~\bibnamefont {Sota}},\ and\ \bibinfo {author} {\bibfnamefont {S.}~\bibnamefont {Yunoki}},\ }\href@noop {} {\bibinfo {title} {Spin and charge dynamics of doped one-dimensional mott insulators}} (\bibinfo {year} {2024}),\ \Eprint {https://arxiv.org/abs/2404.11412} {arXiv:2404.11412 [cond-mat.str-el]} \BibitemShut {NoStop}%
\bibitem [{\citenamefont {Bisogni}\ \emph {et~al.}(2014)\citenamefont {Bisogni}, \citenamefont {Kourtis}, \citenamefont {Monney}, \citenamefont {Zhou}, \citenamefont {Kraus}, \citenamefont {Sekar}, \citenamefont {Strocov}, \citenamefont {B\"uchner}, \citenamefont {van~den Brink}, \citenamefont {Braicovich}, \citenamefont {Schmitt}, \citenamefont {Daghofer},\ and\ \citenamefont {Geck}}]{Bisogni2014}%
  \BibitemOpen
  \bibfield  {author} {\bibinfo {author} {\bibfnamefont {V.}~\bibnamefont {Bisogni}}, \bibinfo {author} {\bibfnamefont {S.}~\bibnamefont {Kourtis}}, \bibinfo {author} {\bibfnamefont {C.}~\bibnamefont {Monney}}, \bibinfo {author} {\bibfnamefont {K.}~\bibnamefont {Zhou}}, \bibinfo {author} {\bibfnamefont {R.}~\bibnamefont {Kraus}}, \bibinfo {author} {\bibfnamefont {C.}~\bibnamefont {Sekar}}, \bibinfo {author} {\bibfnamefont {V.}~\bibnamefont {Strocov}}, \bibinfo {author} {\bibfnamefont {B.}~\bibnamefont {B\"uchner}}, \bibinfo {author} {\bibfnamefont {J.}~\bibnamefont {van~den Brink}}, \bibinfo {author} {\bibfnamefont {L.}~\bibnamefont {Braicovich}}, \bibinfo {author} {\bibfnamefont {T.}~\bibnamefont {Schmitt}}, \bibinfo {author} {\bibfnamefont {M.}~\bibnamefont {Daghofer}},\ and\ \bibinfo {author} {\bibfnamefont {J.}~\bibnamefont {Geck}},\ }\bibfield  {title} {\bibinfo {title} {Femtosecond dynamics of momentum-dependent magnetic excitations from resonant inelastic x-ray scattering in
  ${\mathrm{cacu}}_{2}{\mathrm{o}}_{3}$},\ }\href {https://doi.org/10.1103/PhysRevLett.112.147401} {\bibfield  {journal} {\bibinfo  {journal} {Phys. Rev. Lett.}\ }\textbf {\bibinfo {volume} {112}},\ \bibinfo {pages} {147401} (\bibinfo {year} {2014})}\BibitemShut {NoStop}%
\bibitem [{\citenamefont {Forte}\ \emph {et~al.}(2011)\citenamefont {Forte}, \citenamefont {Cuoco}, \citenamefont {Noce},\ and\ \citenamefont {van~den Brink}}]{Forte2011}%
  \BibitemOpen
  \bibfield  {author} {\bibinfo {author} {\bibfnamefont {F.}~\bibnamefont {Forte}}, \bibinfo {author} {\bibfnamefont {M.}~\bibnamefont {Cuoco}}, \bibinfo {author} {\bibfnamefont {C.}~\bibnamefont {Noce}},\ and\ \bibinfo {author} {\bibfnamefont {J.}~\bibnamefont {van~den Brink}},\ }\bibfield  {title} {\bibinfo {title} {Doping dependence of magnetic excitations of one-dimensional cuprates as probed by resonant inelastic x-ray scattering},\ }\href {https://doi.org/10.1103/PhysRevB.83.245133} {\bibfield  {journal} {\bibinfo  {journal} {Phys. Rev. B}\ }\textbf {\bibinfo {volume} {83}},\ \bibinfo {pages} {245133} (\bibinfo {year} {2011})}\BibitemShut {NoStop}%
\bibitem [{Sup()}]{Sup}%
  \BibitemOpen
  \href@noop {} {}\bibinfo {note} {See supplemental material for additional details.}\BibitemShut {Stop}%
\bibitem [{\citenamefont {Fink}\ \emph {et~al.}(1994)\citenamefont {Fink}, \citenamefont {Nücker}, \citenamefont {Pellegrin}, \citenamefont {Romberg}, \citenamefont {Alexander},\ and\ \citenamefont {Knupfer}}]{Fink_1994}%
  \BibitemOpen
  \bibfield  {author} {\bibinfo {author} {\bibfnamefont {J.}~\bibnamefont {Fink}}, \bibinfo {author} {\bibfnamefont {N.}~\bibnamefont {Nücker}}, \bibinfo {author} {\bibfnamefont {E.}~\bibnamefont {Pellegrin}}, \bibinfo {author} {\bibfnamefont {H.}~\bibnamefont {Romberg}}, \bibinfo {author} {\bibfnamefont {M.}~\bibnamefont {Alexander}},\ and\ \bibinfo {author} {\bibfnamefont {M.}~\bibnamefont {Knupfer}},\ }\bibfield  {title} {\bibinfo {title} {Electron energy-loss and x-ray absorption spectroscopy of cuprate superconductors and related compounds},\ }\href {https://doi.org/10.1016/0368-2048(93)01857-B} {\bibfield  {journal} {\bibinfo  {journal} {Journal of Electron Spectroscopy and Related Phenomena}\ }\textbf {\bibinfo {volume} {66}},\ \bibinfo {pages} {395} (\bibinfo {year} {1994})}\BibitemShut {NoStop}%
\bibitem [{\citenamefont {Neudert}\ \emph {et~al.}(2000)\citenamefont {Neudert}, \citenamefont {Drechsler}, \citenamefont {M\'alek}, \citenamefont {Rosner}, \citenamefont {Kielwein}, \citenamefont {Hu}, \citenamefont {Knupfer}, \citenamefont {Golden}, \citenamefont {Fink}, \citenamefont {N\"ucker}, \citenamefont {Merz}, \citenamefont {Schuppler}, \citenamefont {Motoyama}, \citenamefont {Eisaki}, \citenamefont {Uchida}, \citenamefont {Domke},\ and\ \citenamefont {Kaindl}}]{Neudert2000}%
  \BibitemOpen
  \bibfield  {author} {\bibinfo {author} {\bibfnamefont {R.}~\bibnamefont {Neudert}}, \bibinfo {author} {\bibfnamefont {S.-L.}\ \bibnamefont {Drechsler}}, \bibinfo {author} {\bibfnamefont {J.}~\bibnamefont {M\'alek}}, \bibinfo {author} {\bibfnamefont {H.}~\bibnamefont {Rosner}}, \bibinfo {author} {\bibfnamefont {M.}~\bibnamefont {Kielwein}}, \bibinfo {author} {\bibfnamefont {Z.}~\bibnamefont {Hu}}, \bibinfo {author} {\bibfnamefont {M.}~\bibnamefont {Knupfer}}, \bibinfo {author} {\bibfnamefont {M.~S.}\ \bibnamefont {Golden}}, \bibinfo {author} {\bibfnamefont {J.}~\bibnamefont {Fink}}, \bibinfo {author} {\bibfnamefont {N.}~\bibnamefont {N\"ucker}}, \bibinfo {author} {\bibfnamefont {M.}~\bibnamefont {Merz}}, \bibinfo {author} {\bibfnamefont {S.}~\bibnamefont {Schuppler}}, \bibinfo {author} {\bibfnamefont {N.}~\bibnamefont {Motoyama}}, \bibinfo {author} {\bibfnamefont {H.}~\bibnamefont {Eisaki}}, \bibinfo {author} {\bibfnamefont {S.}~\bibnamefont {Uchida}}, \bibinfo {author} {\bibfnamefont {M.}~\bibnamefont
  {Domke}},\ and\ \bibinfo {author} {\bibfnamefont {G.}~\bibnamefont {Kaindl}},\ }\bibfield  {title} {\bibinfo {title} {{Four-band extended Hubbard Hamiltonian for the one-dimensional cuprate ${\mathrm{Sr}}_{2}{\mathrm{CuO}}_{3}:$ Distribution of oxygen holes and its relation to strong intersite Coulomb interaction}},\ }\href {https://doi.org/10.1103/PhysRevB.62.10752} {\bibfield  {journal} {\bibinfo  {journal} {Phys. Rev. B}\ }\textbf {\bibinfo {volume} {62}},\ \bibinfo {pages} {10752} (\bibinfo {year} {2000})}\BibitemShut {NoStop}%
\bibitem [{\citenamefont {Salluzzo}\ \emph {et~al.}(2007)\citenamefont {Salluzzo}, \citenamefont {Ghiringhelli}, \citenamefont {Brookes}, \citenamefont {De~Luca}, \citenamefont {Fracassi},\ and\ \citenamefont {Vaglio}}]{Salluzzo2007}%
  \BibitemOpen
  \bibfield  {author} {\bibinfo {author} {\bibfnamefont {M.}~\bibnamefont {Salluzzo}}, \bibinfo {author} {\bibfnamefont {G.}~\bibnamefont {Ghiringhelli}}, \bibinfo {author} {\bibfnamefont {N.~B.}\ \bibnamefont {Brookes}}, \bibinfo {author} {\bibfnamefont {G.~M.}\ \bibnamefont {De~Luca}}, \bibinfo {author} {\bibfnamefont {F.}~\bibnamefont {Fracassi}},\ and\ \bibinfo {author} {\bibfnamefont {R.}~\bibnamefont {Vaglio}},\ }\bibfield  {title} {\bibinfo {title} {{Superconducting-insulator transition driven by out-of-plane carrier localization in ${\mathrm{Nd}}_{1.2}{\mathrm{Ba}}_{1.8}{\mathrm{Cu}}_{3}{\mathrm{O}}_{7+x}$ ultrathin films}},\ }\href {https://doi.org/10.1103/PhysRevB.75.054519} {\bibfield  {journal} {\bibinfo  {journal} {Phys. Rev. B}\ }\textbf {\bibinfo {volume} {75}},\ \bibinfo {pages} {054519} (\bibinfo {year} {2007})}\BibitemShut {NoStop}%
\bibitem [{\citenamefont {Chen}\ \emph {et~al.}(1992)\citenamefont {Chen}, \citenamefont {Tjeng}, \citenamefont {Kwo}, \citenamefont {Kao}, \citenamefont {Rudolf}, \citenamefont {Sette},\ and\ \citenamefont {Fleming}}]{Chen_PRL_1992}%
  \BibitemOpen
  \bibfield  {author} {\bibinfo {author} {\bibfnamefont {C.~T.}\ \bibnamefont {Chen}}, \bibinfo {author} {\bibfnamefont {L.~H.}\ \bibnamefont {Tjeng}}, \bibinfo {author} {\bibfnamefont {J.}~\bibnamefont {Kwo}}, \bibinfo {author} {\bibfnamefont {H.~L.}\ \bibnamefont {Kao}}, \bibinfo {author} {\bibfnamefont {P.}~\bibnamefont {Rudolf}}, \bibinfo {author} {\bibfnamefont {F.}~\bibnamefont {Sette}},\ and\ \bibinfo {author} {\bibfnamefont {R.~M.}\ \bibnamefont {Fleming}},\ }\bibfield  {title} {\bibinfo {title} {{Out-of-plane orbital characters of intrinsic and doped holes in ${\mathrm{La}}_{2\mathrm{\ensuremath{-}}\mathit{x}}$${\mathrm{Sr}}_{\mathit{x}}$${\mathrm{CuO}}_{4}$}},\ }\href {https://doi.org/10.1103/PhysRevLett.68.2543} {\bibfield  {journal} {\bibinfo  {journal} {Phys. Rev. Lett.}\ }\textbf {\bibinfo {volume} {68}},\ \bibinfo {pages} {2543} (\bibinfo {year} {1992})}\BibitemShut {NoStop}%
\bibitem [{\citenamefont {van Veenendaal}\ and\ \citenamefont {Sawatzky}(1994)}]{Veenendaal_1994}%
  \BibitemOpen
  \bibfield  {author} {\bibinfo {author} {\bibfnamefont {M.~A.}\ \bibnamefont {van Veenendaal}}\ and\ \bibinfo {author} {\bibfnamefont {G.~A.}\ \bibnamefont {Sawatzky}},\ }\bibfield  {title} {\bibinfo {title} {Intersite interactions in cu l-edge xps, xas, and xes of doped and undoped cu compounds},\ }\href {https://doi.org/10.1103/PhysRevB.49.3473} {\bibfield  {journal} {\bibinfo  {journal} {Phys. Rev. B}\ }\textbf {\bibinfo {volume} {49}},\ \bibinfo {pages} {3473} (\bibinfo {year} {1994})}\BibitemShut {NoStop}%
\bibitem [{\citenamefont {Ghiringhelli}\ \emph {et~al.}(2007)\citenamefont {Ghiringhelli}, \citenamefont {Brookes}, \citenamefont {Dallera}, \citenamefont {Tagliaferri},\ and\ \citenamefont {Braicovich}}]{Ghiringhelli_PRB_2007}%
  \BibitemOpen
  \bibfield  {author} {\bibinfo {author} {\bibfnamefont {G.}~\bibnamefont {Ghiringhelli}}, \bibinfo {author} {\bibfnamefont {N.~B.}\ \bibnamefont {Brookes}}, \bibinfo {author} {\bibfnamefont {C.}~\bibnamefont {Dallera}}, \bibinfo {author} {\bibfnamefont {A.}~\bibnamefont {Tagliaferri}},\ and\ \bibinfo {author} {\bibfnamefont {L.}~\bibnamefont {Braicovich}},\ }\bibfield  {title} {\bibinfo {title} {Sensitivity to hole doping of cu ${L}_{3}$ resonant spectroscopies: Inelastic x-ray scattering and photoemission of $\mathrm{La}_{2\ensuremath{-}x}\mathrm{Sr}_{x}\mathrm{Cu}\mathrm{O}_{4}$},\ }\href {https://doi.org/10.1103/PhysRevB.76.085116} {\bibfield  {journal} {\bibinfo  {journal} {Phys. Rev. B}\ }\textbf {\bibinfo {volume} {76}},\ \bibinfo {pages} {085116} (\bibinfo {year} {2007})}\BibitemShut {NoStop}%
\bibitem [{\citenamefont {Wang}\ \emph {et~al.}(2014)\citenamefont {Wang}, \citenamefont {Liang}, \citenamefont {Liu}, \citenamefont {Huang}, \citenamefont {Yang}, \citenamefont {Luo}, \citenamefont {Yang}, \citenamefont {Hu}, \citenamefont {Jin},\ and\ \citenamefont {Gao}}]{Wang_jes_2014}%
  \BibitemOpen
  \bibfield  {author} {\bibinfo {author} {\bibfnamefont {H.}~\bibnamefont {Wang}}, \bibinfo {author} {\bibfnamefont {W.}~\bibnamefont {Liang}}, \bibinfo {author} {\bibfnamefont {Q.}~\bibnamefont {Liu}}, \bibinfo {author} {\bibfnamefont {H.}~\bibnamefont {Huang}}, \bibinfo {author} {\bibfnamefont {M.}~\bibnamefont {Yang}}, \bibinfo {author} {\bibfnamefont {Z.}~\bibnamefont {Luo}}, \bibinfo {author} {\bibfnamefont {Y.}~\bibnamefont {Yang}}, \bibinfo {author} {\bibfnamefont {S.}~\bibnamefont {Hu}}, \bibinfo {author} {\bibfnamefont {C.}~\bibnamefont {Jin}},\ and\ \bibinfo {author} {\bibfnamefont {C.}~\bibnamefont {Gao}},\ }\bibfield  {title} {\bibinfo {title} {{Polarization-dependent soft X-ray absorption of over-doped superconducting Sr2CuO$_{3+\delta}$ single crystal}},\ }\href {https://doi.org/10.1016/j.elspec.2014.01.001} {\bibfield  {journal} {\bibinfo  {journal} {Journal of Electron Spectroscopy and Related Phenomena}\ }\textbf {\bibinfo {volume} {196}},\ \bibinfo {pages} {61} (\bibinfo {year}
  {2014})}\BibitemShut {NoStop}%
\bibitem [{\citenamefont {Nikitin}\ \emph {et~al.}(2021)\citenamefont {Nikitin}, \citenamefont {Nishimoto}, \citenamefont {Fan}, \citenamefont {Wu}, \citenamefont {Wu}, \citenamefont {Sukhanov}, \citenamefont {Brando}, \citenamefont {Pavlovskii}, \citenamefont {Xu}, \citenamefont {Vasylechko}, \citenamefont {Yu},\ and\ \citenamefont {Podlesnyak}}]{Nikitin2021}%
  \BibitemOpen
  \bibfield  {author} {\bibinfo {author} {\bibfnamefont {S.~E.}\ \bibnamefont {Nikitin}}, \bibinfo {author} {\bibfnamefont {S.}~\bibnamefont {Nishimoto}}, \bibinfo {author} {\bibfnamefont {Y.}~\bibnamefont {Fan}}, \bibinfo {author} {\bibfnamefont {J.}~\bibnamefont {Wu}}, \bibinfo {author} {\bibfnamefont {L.~S.}\ \bibnamefont {Wu}}, \bibinfo {author} {\bibfnamefont {A.~S.}\ \bibnamefont {Sukhanov}}, \bibinfo {author} {\bibfnamefont {M.}~\bibnamefont {Brando}}, \bibinfo {author} {\bibfnamefont {N.~S.}\ \bibnamefont {Pavlovskii}}, \bibinfo {author} {\bibfnamefont {J.}~\bibnamefont {Xu}}, \bibinfo {author} {\bibfnamefont {L.}~\bibnamefont {Vasylechko}}, \bibinfo {author} {\bibfnamefont {R.}~\bibnamefont {Yu}},\ and\ \bibinfo {author} {\bibfnamefont {A.}~\bibnamefont {Podlesnyak}},\ }\bibfield  {title} {\bibinfo {title} {{Multiple fermion scattering in the weakly coupled spin-chain compound YbAlO$_3$}},\ }\bibfield  {journal} {\bibinfo  {journal} {Nature Communications}\ }\textbf {\bibinfo {volume} {12}},\ \href
  {https://doi.org/10.1038/s41467-021-23585-z} {10.1038/s41467-021-23585-z} (\bibinfo {year} {2021})\BibitemShut {NoStop}%
\bibitem [{\citenamefont {Wang}\ \emph {et~al.}(2021)\citenamefont {Wang}, \citenamefont {Chen}, \citenamefont {Shi}, \citenamefont {Moritz}, \citenamefont {Shen},\ and\ \citenamefont {Devereaux}}]{Wang_2021}%
  \BibitemOpen
  \bibfield  {author} {\bibinfo {author} {\bibfnamefont {Y.}~\bibnamefont {Wang}}, \bibinfo {author} {\bibfnamefont {Z.}~\bibnamefont {Chen}}, \bibinfo {author} {\bibfnamefont {T.}~\bibnamefont {Shi}}, \bibinfo {author} {\bibfnamefont {B.}~\bibnamefont {Moritz}}, \bibinfo {author} {\bibfnamefont {Z.-X.}\ \bibnamefont {Shen}},\ and\ \bibinfo {author} {\bibfnamefont {T.~P.}\ \bibnamefont {Devereaux}},\ }\bibfield  {title} {\bibinfo {title} {Phonon-mediated long-range attractive interaction in one-dimensional cuprates},\ }\href {https://doi.org/10.1103/PhysRevLett.127.197003} {\bibfield  {journal} {\bibinfo  {journal} {Phys. Rev. Lett.}\ }\textbf {\bibinfo {volume} {127}},\ \bibinfo {pages} {197003} (\bibinfo {year} {2021})}\BibitemShut {NoStop}%
\bibitem [{\citenamefont {Tang}\ \emph {et~al.}(2023)\citenamefont {Tang}, \citenamefont {Moritz}, \citenamefont {Peng}, \citenamefont {Shen},\ and\ \citenamefont {Devereaux}}]{Tang2023}%
  \BibitemOpen
  \bibfield  {author} {\bibinfo {author} {\bibfnamefont {T.}~\bibnamefont {Tang}}, \bibinfo {author} {\bibfnamefont {B.}~\bibnamefont {Moritz}}, \bibinfo {author} {\bibfnamefont {C.}~\bibnamefont {Peng}}, \bibinfo {author} {\bibfnamefont {Z.~X.}\ \bibnamefont {Shen}},\ and\ \bibinfo {author} {\bibfnamefont {T.~P.}\ \bibnamefont {Devereaux}},\ }\bibfield  {title} {\bibinfo {title} {Traces of electron-phonon coupling in one-dimensional cuprates},\ }\bibfield  {journal} {\bibinfo  {journal} {Nature Communications}\ }\textbf {\bibinfo {volume} {14}},\ \href {https://doi.org/10.1038/s41467-023-38408-6} {10.1038/s41467-023-38408-6} (\bibinfo {year} {2023})\BibitemShut {NoStop}%
\bibitem [{\citenamefont {Wang}\ \emph {et~al.}(2024)\citenamefont {Wang}, \citenamefont {Wu}, \citenamefont {Jiang},\ and\ \citenamefont {Yao}}]{Wang2024}%
  \BibitemOpen
  \bibfield  {author} {\bibinfo {author} {\bibfnamefont {H.-X.}\ \bibnamefont {Wang}}, \bibinfo {author} {\bibfnamefont {Y.-M.}\ \bibnamefont {Wu}}, \bibinfo {author} {\bibfnamefont {Y.-F.}\ \bibnamefont {Jiang}},\ and\ \bibinfo {author} {\bibfnamefont {H.}~\bibnamefont {Yao}},\ }\bibfield  {title} {\bibinfo {title} {Spectral properties of a one-dimensional extended hubbard model from bosonization and time-dependent variational principle: Applications to one-dimensional cuprates},\ }\href {https://doi.org/10.1103/PhysRevB.109.045102} {\bibfield  {journal} {\bibinfo  {journal} {Phys. Rev. B}\ }\textbf {\bibinfo {volume} {109}},\ \bibinfo {pages} {045102} (\bibinfo {year} {2024})}\BibitemShut {NoStop}%
\bibitem [{\citenamefont {Devereaux}\ \emph {et~al.}(2016)\citenamefont {Devereaux}, \citenamefont {Shvaika}, \citenamefont {Wu}, \citenamefont {Wohlfeld}, \citenamefont {Jia}, \citenamefont {Wang}, \citenamefont {Moritz}, \citenamefont {Chaix}, \citenamefont {Lee}, \citenamefont {Shen}, \citenamefont {Ghiringhelli},\ and\ \citenamefont {Braicovich}}]{Devereaux2016}%
  \BibitemOpen
  \bibfield  {author} {\bibinfo {author} {\bibfnamefont {T.~P.}\ \bibnamefont {Devereaux}}, \bibinfo {author} {\bibfnamefont {A.~M.}\ \bibnamefont {Shvaika}}, \bibinfo {author} {\bibfnamefont {K.}~\bibnamefont {Wu}}, \bibinfo {author} {\bibfnamefont {K.}~\bibnamefont {Wohlfeld}}, \bibinfo {author} {\bibfnamefont {C.~J.}\ \bibnamefont {Jia}}, \bibinfo {author} {\bibfnamefont {Y.}~\bibnamefont {Wang}}, \bibinfo {author} {\bibfnamefont {B.}~\bibnamefont {Moritz}}, \bibinfo {author} {\bibfnamefont {L.}~\bibnamefont {Chaix}}, \bibinfo {author} {\bibfnamefont {W.-S.}\ \bibnamefont {Lee}}, \bibinfo {author} {\bibfnamefont {Z.-X.}\ \bibnamefont {Shen}}, \bibinfo {author} {\bibfnamefont {G.}~\bibnamefont {Ghiringhelli}},\ and\ \bibinfo {author} {\bibfnamefont {L.}~\bibnamefont {Braicovich}},\ }\bibfield  {title} {\bibinfo {title} {Directly characterizing the relative strength and momentum dependence of electron-phonon coupling using resonant inelastic x-ray scattering},\ }\href
  {https://doi.org/10.1103/PhysRevX.6.041019} {\bibfield  {journal} {\bibinfo  {journal} {Phys. Rev. X}\ }\textbf {\bibinfo {volume} {6}},\ \bibinfo {pages} {041019} (\bibinfo {year} {2016})}\BibitemShut {NoStop}%
\bibitem [{\citenamefont {Jost}\ \emph {et~al.}(2024)\citenamefont {Jost}, \citenamefont {Huang}, \citenamefont {Rossi}, \citenamefont {Singh}, \citenamefont {Huang}, \citenamefont {Lee}, \citenamefont {Zheng}, \citenamefont {Mitchell}, \citenamefont {Moritz}, \citenamefont {Shen}, \citenamefont {Devereaux},\ and\ \citenamefont {Lee}}]{Jost2024}%
  \BibitemOpen
  \bibfield  {author} {\bibinfo {author} {\bibfnamefont {D.}~\bibnamefont {Jost}}, \bibinfo {author} {\bibfnamefont {H.-Y.}\ \bibnamefont {Huang}}, \bibinfo {author} {\bibfnamefont {M.}~\bibnamefont {Rossi}}, \bibinfo {author} {\bibfnamefont {A.}~\bibnamefont {Singh}}, \bibinfo {author} {\bibfnamefont {D.-J.}\ \bibnamefont {Huang}}, \bibinfo {author} {\bibfnamefont {Y.}~\bibnamefont {Lee}}, \bibinfo {author} {\bibfnamefont {H.}~\bibnamefont {Zheng}}, \bibinfo {author} {\bibfnamefont {J.~F.}\ \bibnamefont {Mitchell}}, \bibinfo {author} {\bibfnamefont {B.}~\bibnamefont {Moritz}}, \bibinfo {author} {\bibfnamefont {Z.-X.}\ \bibnamefont {Shen}}, \bibinfo {author} {\bibfnamefont {T.~P.}\ \bibnamefont {Devereaux}},\ and\ \bibinfo {author} {\bibfnamefont {W.-S.}\ \bibnamefont {Lee}},\ }\bibfield  {title} {\bibinfo {title} {Low temperature dynamic polaron liquid in a manganite exhibiting colossal magnetoresistance},\ }\href {https://doi.org/10.1103/PhysRevLett.132.186502} {\bibfield  {journal} {\bibinfo  {journal}
  {Phys. Rev. Lett.}\ }\textbf {\bibinfo {volume} {132}},\ \bibinfo {pages} {186502} (\bibinfo {year} {2024})}\BibitemShut {NoStop}%
\end{thebibliography}%

\end{document}